\documentclass[twocolumn]{aastex62}
\usepackage[T1]{fontenc}
\usepackage{amsmath,amssymb}
\usepackage{epstopdf}
\usepackage{ulem}
\usepackage{multirow}
\usepackage{todonotes}
\usepackage{lineno}

\usepackage{soul}
\usepackage{float}
\usepackage{xcolor}

\newcommand{\ab}{\textcolor{black}}

\shorttitle{Electron Reflection at Oblique Shocks}
\shortauthors{Morris et al.}

\def\ee{\end{equation}}
\def\be{\begin{equation}}

\newcommand{\thbn}{\theta_\mathrm{Bn}}

\newcommand{\ompe}{\omega_\mathrm{pe}}

\newcommand{\mi}{m_\mathrm{i}}

\newcommand{\lse}{\lambda_\mathrm{se}}

\newcommand{\ppar}{p_{\parallel}}
\newcommand{\pperp}{p_{\bot}}

\begin{document}

\title{Pre-acceleration in the Electron Foreshock I: Electron Acoustic Waves}

\correspondingauthor{Paul J. Morris}
\email{paul.morris@desy.de}

\author[0000-0002-8533-8232]{Paul J. Morris}
\affil{Deutsches Elektronen-Synchrotron DESY, Platanenallee 6, 15738 Zeuthen, Germany}

\author[0000-0002-5680-0766]{Artem Bohdan}
\affil{Deutsches Elektronen-Synchrotron DESY, Platanenallee 6, 15738 Zeuthen, Germany}

\author[0000-0002-3440-3225]{Martin~S.~Weidl}
\affiliation{Max-Planck-Institut für Plasmaphysik, Boltzmannstr. 2, DE-85748 Garching, Germany}

\author[0000-0001-7861-1707]{Martin Pohl}
\affil{Deutsches Elektronen-Synchrotron DESY, Platanenallee 6, 15738 Zeuthen, Germany}
\affil{Institute of Physics and Astronomy, University of Potsdam, D-14476 Potsdam, Germany}

\begin{abstract}

To undergo diffusive shock acceleration, electrons need to be pre-accelerated to increase their energies by several  
orders of magnitude, else their gyro-radii are smaller than the finite width of the shock. In oblique shocks, where the upstream magnetic field orientation is neither parallel or perpendicular to the shock normal, electrons can escape to the shock upstream, modifying the shock foot to a region called the electron foreshock. To determine the pre-acceleration in this region, we undertake PIC simulations of oblique shocks while varying the obliquity and in-plane angles. We show that while the proportion of reflected electrons is negligible for $\thbn = 74.3^\circ$, it increases to $R \sim 5\%$ for $\thbn = 30^\circ$, and that, via the electron acoustic instability, these electrons power electrostatic waves upstream with energy density proportional to $R^{0.6}$
and a wavelength $\approx 2 \lambda_{\rm{se}}$, where $\lambda_{\rm{se}}$ is the electron skin length. While the initial reflection mechanism is typically a combination of shock surfing acceleration and magnetic mirroring, we show that once the electrostatic waves have been generated upstream they themselves can increase the momenta of upstream electrons parallel to the magnetic field. In $\lesssim 1\%$ of cases, upstream electrons are prematurely turned away from the shock and never injected downstream. In contrast, a similar fraction are re-scattered back towards the shock after reflection, re-interact with the shock with energies much greater than thermal, and cross into the downstream.

\end{abstract}


\section{Introduction}\label{introduction}

Astrophysical shocks occur ubiquitously in our universe, with detections of non-thermal X-ray emission from sources such as supernova remnants implying the presence of accelerated electrons. Their primary acceleration mechanism is widely accepted to be Diffusive shock acceleration (DSA) \citep{Bell1978a}, yet alone it does not provide a complete picture. This is because of the injection problem: for DSA to work, the Larmor radii of the particles it accelerates must be comparable to the finite shock width, a criterion not met by thermal electrons. Therefore, they require some pre-acceleration, the exact physical nature of which is crucial to fully comprehend many astrophysical phenomena. 

The study of electron-scale phenomena, which may be responsible for pre-acceleration, requires a method capable of resolving these small kinectic scales. Particle-in-cell (PIC) simulations fulfill this criterion and have previously proved invaluable in helping to establish the contribution to electron pre-acceleration in perpendicular shocks from mechanisms such as shock surfing acceleration, magnetic reconnection, and stochastic Fermi acceleration \citep{Bohdan2017,Bohdan2019a,Bohdan2019b,Bohdan2020a,Bohdan2020b}. 

The obliquity angle, $\thbn$, defined as the angle subtended between the shock normal and upstream magnetic field, can heavily influence the behavior of the shock, with DSA the dominant particle energization process for quasi-parallel shocks and Shock-drift acceleration increasingly important at larger subluminal magnetic field inclination angles \citep{Ellison1995,Sironi2009}. In comparison to perpendicular shocks (where $\thbn=90^\circ$), which typically have a well-defined shock transition with a narrow foot region, oblique shocks permit the escape (or reflection) of particles along the upstream magnetic field lines and back into the shock upstream, which creates extended regions known as the electron and ion foreshocks, depending on the particle species. \citep[e.g.][]{Burgess1995,Fitzenreiter1995,Treumann2009}. 

In previous studies of oblique shocks, the mechanism of electron reflection to the upstream region has been shown to be mirror reflection \citep{2005MNRAS.362..833H}. More energetic electrons are favorably reflected back upstream, with these electrons usually pre-energized by shock-surfing acceleration \citep{Amano2007}. Notably, 1D PIC simulations have demonstrated that such reflected electrons are capable of generating electrostatic waves in the shock foot, where they may undergo shock drift acceleration (SDA) before being injected downstream \citep{Xu2020,2021ApJ...921L..14K}.  

As any electrons upstream of an oblique shock must first encounter the foreshock and any turbulence present there before interacting with the shock, understanding the properties of this region and how they influence the upstream plasma are of utmost importance when addressing electron pre-acceleration and the shock injection problem.

In this paper, we investigate the impact of changing the orientation of the upstream magnetic field on the electron foreshock. The structure of the manuscript is as follows: first, we outline the initial mechanism responsible for reflecting electrons. Then, we show that these reflected electrons generate electron acoustic waves, and examine their properties. Finally, we look at what happens to the energy associated with these waves, and how they effect electrons in the upstream plasma.



\section{Simulation Setup} \label{sec:setup}

\begin{figure}
    \centering
    \includegraphics[width=\columnwidth]{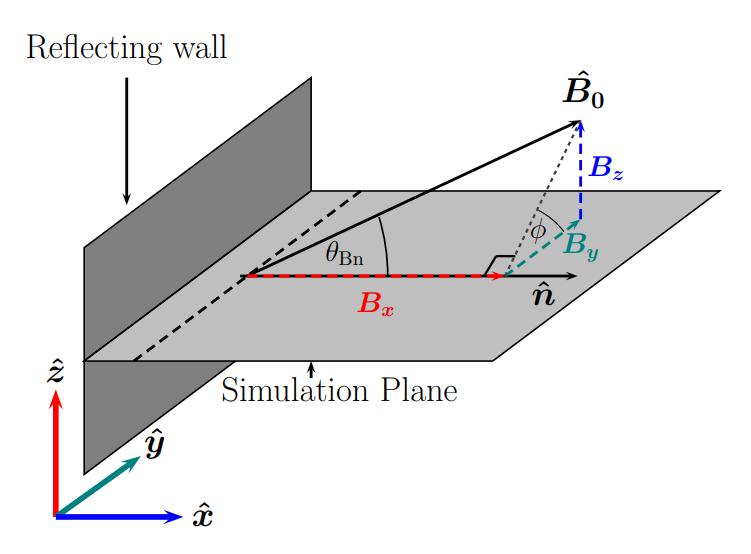}
    \caption{Reflecting wall setup of our simulations. Here, $\phi$ is the angle $\boldsymbol{{B_0}}$ makes relative to the simulation plane. The angle $\thbn$ is defined as that between the shock normal, $\boldsymbol{\hat{n}}$, and the magnetic field vector, $\boldsymbol{{B_0}}$ such that \mbox{$\boldsymbol{B_0} = B_0( \cos \theta_{\rm{Bn}},  \sin \theta_{\rm{Bn}} \cos \phi, \sin \theta_{\rm{Bn}} \sin \phi)$}. }
    \label{fig:simSetup}
\end{figure}

The work presented here makes use of code adapted from TRISTAN \citep{Buneman1993} which allows us to study fully relativistic shocks while tracking 2-spatial and all 3 velocity components (2D3V) of the individual particles in our plasma \citep{Niemiec2008}. Particle positions are updated using the Vay solver \citep{Vay2008}.  When applied to perpendicular shocks, the data from these simulations have proven to accurately reproduce the fundamental physical processes present in larger and computationally more expensive full 3D simulations \citep[e.g.][]{Bohdan2017,Matsumoto2017} and measured in-situ \citep{Bohdan2020b,Bohdan2021}. This version of the code also permits the tracing of individual particles, where we can track the components of their velocities and local fields, allowing for a detailed analysis of the underlying physical processes. 

Previous works investigating perpendicular shocks successfully employed a flow-flow method in which two counter-streaming plasma slabs collide and generate two shocks \citep{Wieland2016,Bohdan2017,Bohdan2019a,Bohdan2019b,Bohdan2020a,Bohdan2020b,Bohdan2021}. Relative to perpendicular shocks, an investigation of an oblique shock requires a larger simulation box, and therefore a greater computational expense, to ensure reflected electrons are captured, so in this work we instead implement a reflecting wall setup (injection method). In this configuration, a single shock is produced by reflecting an incoming plasma beam from a conducting wall located at $x=0$ in our simulation box, where the velocity transformation $v_x \rightarrow -v_x$ is performed ($v_y$ and $v_z$ remain unaltered) \citep{Quest1985,Burgess1989}. To prevent an electromagnetic transient from being induced by the large values of $\nabla \times \boldsymbol{E}$ and the corresponding $\partial \boldsymbol{B} / \partial t$, we taper the fields near the reflection boundary following \citet{Wieland2016}, where a drift velocity is also assigned to ions in these regions to compensate the now non-zero $\nabla \times \boldsymbol{B}$. After reflection, this drift velocity is removed from the ions because the drifts at a real contact discontinuity compensate each other. By adding further domains, fresh plasma is regularly injected into the upstream from the right boundary of the simulation box with the same initial parameters as outlined above. One can visualize the injection of new particles into a new domain as dividing it into small sub-regions with one of each particle species injected at a random, but identical, position within each sub-region. The sizes of these sub-regions are  selected such that the desired number of particles-per-cell, $n_{\rm ppc}$ is returned.

Excepting the upstream magnetic field configuration, all of the simulations presented here use the same initial parameters, which are expressed in Heaviside--Lorentz units. First, we initialize upstream plasma with a bulk velocity of $\boldsymbol{v_{\rm ups}} = -0.2c~ \boldsymbol{\hat{x}}$ in the simulation plane, which corresponds to a shock velocity of $v_{\rm{sh}} \simeq 0.264c$ in the upstream rest frame (or $v_{\rm{sh}}' \simeq 0.0673c$ in the simulation frame). This upstream plasma is initialized with a uniform large-scale magnetic field with a field orientation described by the obliquity angle, $\theta_{\rm{Bn}}$ and angle relative to the simulation plane, $\phi$, such that \mbox{$\boldsymbol{B_0} = B_0( \cos \theta_{\rm{Bn}},  \sin \theta_{\rm{Bn}} \cos \phi, \sin \theta_{\rm{Bn}} \sin \phi)$}. This gives rise to a motional electric field, which is computed as $\boldsymbol{E} = - \boldsymbol{v_{\rm ups}} \times \boldsymbol{B_0}$. Such a magnetic field configuration results in particles which effectively have three degrees of freedom, thus the adiabatic index is $\gamma = 5/3$. The shock compression ratio depends on this quantity, and for all simulations here has the value \ab{$r \approx 4$.}  

We assume that upstream ions and electrons are initially in thermal equilibrium with each other, such that $k_B T_i = k_B T_e = m_i v_{i, th}^2/2 = m_e v_{e, th}^2/2$, where $k_B$ is the Boltzmann constant, $T$ denotes temperature, and $v_{th}$ is the most probable particle speed in the upstream reference frame. The subscripts $i$ and $e$ represent ions and electrons, respectively, with their masses characterized by $m_i/m_e = 64$. This corresponds to numerical values of $v_{th,e} = 0.0713c$ and $v_{th,i} = 0.00089c$ and allow us to determine the sound speed as $c_{\rm s}=(\gamma k_BT_{\rm i}/\mi)^{1/2} = 0.00814c$. These values give the ratio of electron plasma to magnetic pressures (plasma beta) as $\beta =  1$.

With $n_\mathrm{ppc} = n_i = n_e=20$ particles per cell as the number densities of ions and electrons, respectively,
the Alfv\'en velocity is defined as $v_{\rm A}=B_0 c /\sqrt{n_{\rm{ppc}}(m_e+m_i)}$ $=0.0125c$, where $\mu_0$ is the permeability of free space. This sets the Alfv\'enic Mach number as $M_A = v_{\rm{sh}}/v_A \approx 21$, which we define by convention in the upstream reference frame.

Our simulations all share a ratio of electron plasma frequency, $\omega_{pe} = \sqrt{q_e^2 n_e/(\epsilon_0 m_e)}$, to electron cyclotron frequency, $\Omega_e = q_e B_0/m_e$, of $\omega_{pe}/\Omega_e = 9.9$, where $\epsilon_0$ is the vacuum permittivity. The simulation run time is characterized by the ion cyclotron time, $\Omega_{ci}^{-1} = m_i/(q_i B_0)$, where each simulation is run for a total of $T_{\rm tot} = 7.8 \Omega_{ci}^{-1}$ to enable a comparison with the results of \citet{Matsumoto2017}. To adequately resolve all of the important timescales in our simulation, we use a simulation timestep of $\delta t=1/40\,\omega_{\rm pe}^{-1}$.

In these simulations, we study the rates and implications of the reflection of incident electrons at the shock. Accordingly, the transverse size of our simulation box must be large enough to encapsulate electron scale phenomena. From our setup, the gyroradius of thermal electrons measures $r_{g,e} = v_{e,th}/\Omega_{ce} = 14\Delta$, where $\Delta$ represents one cell in our simulation grid, while the electron skin length measures $\lambda_e = c/\omega_{pe} = 20\Delta$. For ions, these quantities are both higher by a factor of $\sqrt{m_i/m_e}$. The transverse size of our simulation box measures 768 $\Delta$ (about $ 55 r_{ge}$ or $ 4.8 \lambda_{i}$). This width is therefore large enough to probe electron physics, yet small enough to save computational expense and allow resources to be allocated to the necessary extension of our simulation box with respect to time so that it contains the reflected electrons. 

In this paper, we present a number of simulations of varying obliquity angles, $\thbn$, between the shock normal and upstream magnetic field. In some cases we also varied the field inclination to the simulation plane, $\phi$. For clarity, these angles are defined in Fig. \ref{fig:simSetup}, with  a full list of used angles provided in Table \ref{tab:sims}. We fix all of the other parameters outlined above, in order to allow a physical comparison of the electron foreshock region in each case. The inclusion of $\theta_{\rm{Bn}}=74.3^{\circ}$ allows us to directly compare our results with full 3D simulations of \citet{Matsumoto2017}, compared to which our parameters are identical. To investigate the effect of the in-plane angle $\phi$, and to allow are more rigorous comparison with the 3D case, we run further simulations with $\phi=90^{\circ}$ for our two largest values of $\theta_{\rm{Bn}}$. For these parameters, the angle at which the shock becomes superluminal is given by $\theta_{\rm{crit}} = \tan^{-1}(c/v_{\rm{sh}}) \sim 75.2^\circ$, thus our range of angles allows a substantial investigation into how the properties of the electron foreshock depend on the initial upstream magnetic field configuration.

\begin{table}
\begin{center}
\begin{tabular}{ cccccccc}
 \hline
 \hline
 Angle & A & B & C & D & E & F & G  \\ 
 \hline
 $\thbn$ & 30 & 45 & 63 & 63 & 63 & 74.3 & 74.3 \\ 
 $\phi$  &  0 &  0 &  0 & 45 & 90 & 0 & 90 \\ 
 \hline
\end{tabular}
 \caption{Table listing the obliquity angle, $\thbn$, and inclination angle, $\phi$, for all simulations presented here. These angles determine the configuration of the upstream magnetic field. All simulations have $M_A \approx 21$.}
 \label{tab:sims}
\end{center}
\end{table}

\section{Shock Structure}

\begin{figure*}
    \centering
    \includegraphics[width=\textwidth]{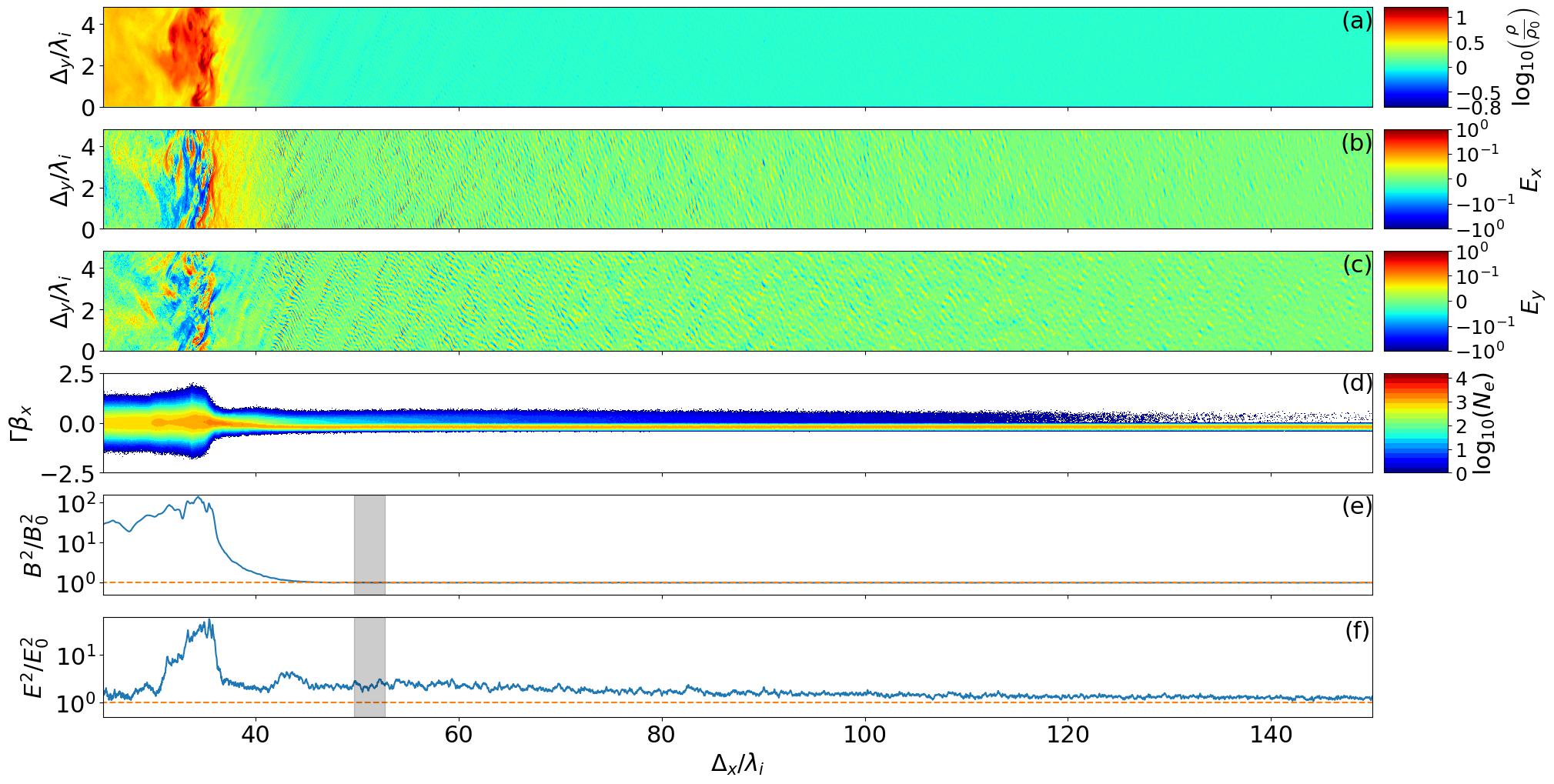}
    \caption{For the simulation with $\thbn = 45^{\circ}$, we show for $t\,\Omega_{ci} = 7.8$ the (a) density map, (b) $E_x$ component of the electric field, (c) $E_y$ component of the electric field, (d) distribution of electron momentum in the $x$-direction, and (e) and (f) the magnetic and electric field profiles, respectively. Note in panels (b) and (c) the presence of electrostatic field in the region with significant electron reflection. The dashed lines in panels (e) and (f) indicate the fields at their far-upstream values. The grey regions in panels (e) and (f) show the region spectra were calculated over.
    }
    \label{fig:shockStructure}
\end{figure*}

Fig. \ref{fig:shockStructure} shows the shock structure for the in-plane simulation with $\thbn = 45^{\circ}$ at the final timestep of the simulation, where $t = 7.8 \Omega_{ci}^{-1}$. Here the shock is fully formed with an overshoot at $x \sim 35 \lambda_i$, and panel (a) shows a map of the electron density relative to the upstream density. In the downstream region towards the left of the figure, we measure a compression ratio close to the analytically expected value of $\rho/\rho_0 \approx 3.97$, with the same true of the magnetic field (panel (e)). From this panel, we also see a smooth transition of the magnetic field profile from downstream to upstream, where it remains at around the initial upstream value where $(B/B_0)^2 = 1$.

In contrast to the magnetic field, panel (f) shows that the electric field steadily decreases with increasing distance from the shock. Panels (b) and (c) show that waves are present in the upstream region where $(E/E_0)^2 > 1$. These waves have no magnetic component and are purely electrostatic, and in this in-plane simulation lie in the $xy$ plane by definition. Visually, we note that a higher density of high amplitude waves occupies the regions with greater $(E/E_0)^2$ (panel (f)), and that the physical length scales on which they occur is of order $\lse$ and comparable to electron-scales, immediately implying that they may be related to electrons. Panel (d) shows the phase-space momentum distribution for all electrons in the selected region of the simulation. There is a clear excess of electrons with positive $\Gamma \beta_x$ with a number density that decreases further from the shock. The regions in which these electrons are present correspond to the region occupied by electrostatic waves, the nature of which we discuss in the following section. Electrons reflected at the shock are able to carry energy back upstream, and understanding what happens to this energy is the object of this paper.

\section{Electron Acoustic Waves} 

Our oblique shock simulations reveal the presence of electrostatic waves which are spatially coincident with the reflected electrons. To determine their nature, we note that following \citet{Bohdan:2022}, their wavelength and growth rate indicate that they are excited by the electron-acoustic instability. In contrast to the Buneman instability \citep{Buneman1958}, which is driven directly at the shock by the ion/electron interaction, the electron-acoustic instability results from the velocity difference between the cold incoming and the hot reflected electrons along the magnetic-field direction \citep{Gary1987}. Simulations with periodic boundary conditions presented in \citet{Bohdan:2022} confirm that reflected ions play no role in generating these electrostatic waves, so we can immediately rule out the electron cyclotron drift instability (EDCI), which is excited by reflected ions interacting with incoming electrons \citep[e.g.][]{Muschietti2006,2020ApJ...900L..24Y}, as the instability that drives them. Similarly, we can rule out the synchrotron maser instability \citep{1991PhFlB...3..818H} as we do not see the characteristic electron ring structure in their phase space distributions  \citep[e.g.][]{2019MNRAS.485.3816P,2020MNRAS.499.2884B}.

As they are driven by reflected electrons, these electrostatic waves are predominantly propagating along the background magnetic field. Since the mechanism driving this electron/electron instability does not involve the more inertial ions, it grows much faster and saturates already far ahead of the shock front. We refer to these waves as electron acoustic waves (EAWs) and discuss their properties, and those of the reflected electrons that drive them, before outlining the electron reflection mechanism in the following sections.

\subsection{Electron Spectra}

To illustrate the properties of the reflected electrons, we calculate spectra and reflection rates over a region of the simulation box. To define this region, we first estimate the location of the shock foot, $x_{\rm{foot}}$, given as the location where the mean ion density averaged across the transverse direction of the simulation box first reaches $\rho/\rho_0 = 1.05$, as measured from upstream to downstream. By defining $x_{\rm{foot}}$ based on ion density we ensure that it remains independent from the electron-induced electrostatic waves that we analyze in this work. It is well established that electrons can undergo shock surfing acceleration along the leading edge of the shock \citep{Bohdan2019a}, after which they can either enter the downstream or be reflected back upstream \citep{Amano2007}. We wish to omit these electrons from our spectra and reflection rate calculations, hence they are calculated in the simulation frame for a region where the location in the simulation box,  $x$,  satisfies $x_{\rm{foot}}(t) +2 \lambda_{si} < x \leq x_{\rm{foot}}(t) + 5 \lambda_{si}$, where $\lambda_{si}$ is the ion skin depth which is resolved by 160 cells. This range is further from the shock than shock surfing electrons and so ensures they are not included in our calculations. Similarly, if this region is too large, we risk including regions of the simulation box which contain no reflected electrons because they have not yet had time to travel this far. This region is shown for $\Omega_{\rm ci}t = 7.8$ for $\thbn = 45^{\circ}$ in panels (e) and (f) of Fig. \ref{fig:shockStructure}.

\begin{figure}
    \includegraphics[width=\columnwidth]{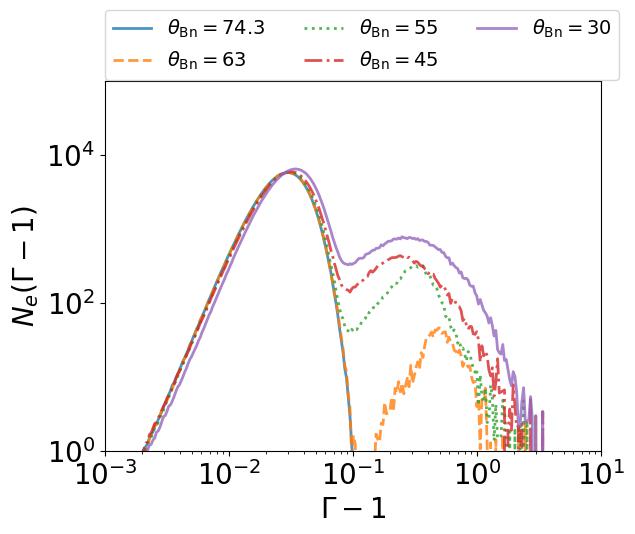}
    \caption{Electron spectra for simulations with $\phi=0^{\circ}$ and varying $\thbn$ as measured in the simulation frame after $\Omega_{ci}t=7.8$. The left and right peaks represent the upstream and reflected electrons, respectively. Spectra were computed for the region between two and five ion skin lengths,  $\lambda_{si}$, upstream of the location, $x_{\rm{foot}}$, where $\rho/\rho_0 = 1.05$, and averaged over the transverse direction of the simulation box.
    } \label{fig:specTheta}
\end{figure}

\subsubsection{In-plane Simulations}

Electron spectra for in-plane simulations after $\Omega_{\rm{ci}}t = 7.8$ are shown in Fig. \ref{fig:specTheta}. Two features immediately stand out. Firstly, we see that for lower values of $\thbn$, more electrons are reflected. The second feature is that the peak value of $\Gamma - 1$ for these reflected electrons becomes higher for larger $\thbn$. The physical interpretation of these features is interconnected. Previous results have verified that upstream electrons arriving at the shock may only be successfully reflected should their velocity, $v_e$, satisfy $v_{e} \cos \thbn \geq v_{\rm{sh}}$, where all quantities are in the shock-frame \citep{Amano2007,Matsumoto2017}. As reflected electrons travel on average in the direction of the upstream magnetic field lines, this result is somewhat intuitive as they require a significant velocity component in the direction parallel to the shock propagation; otherwise they cannot outrun it.

It follows that electrons require larger velocities for larger values of $\thbn$. Furthermore, as less energetic electrons are relatively more abundant due to the lower amount of work done on them to achieve such energies, more of them fulfill the reflection criterion when $\thbn$ is smaller, which in turn leads to a higher reflection rate. Similarly, at larger $\thbn$ the number of electrons eligible for reflection is reduced, hence we measure a smaller reflection rate overall. Fig. \ref{fig:specTheta} shows that the peak value of $\Gamma - 1$ typically increases with $\thbn$, which further supports this explanation. 

\begin{figure}
    \includegraphics[width=0.9\columnwidth]{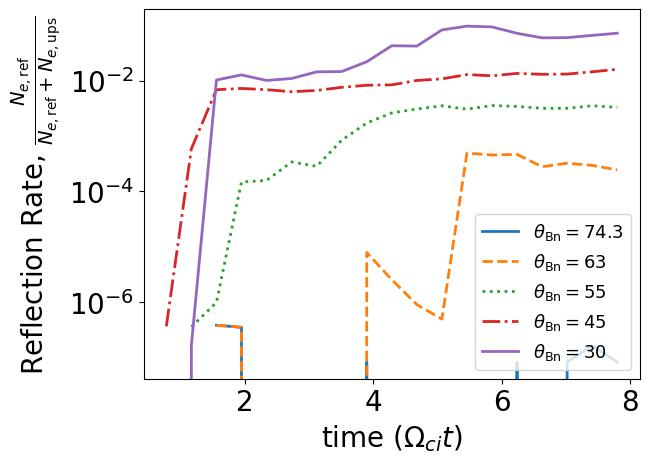}
    \caption{Electron reflection rates for simulations with differing $\thbn$ as function of time in units of ion gyroperiods. When $\thbn$ is smaller, the reflection rate is higher which leads to a more extended electron foreshock. 
    }
    \label{fig:refFrac}
\end{figure}

One important quantity we assess here is the electron reflection rate, which we define as $R = N_{e,\rm{ref}}/(N_{e,\rm{ref}}+N_{e,\rm{ups}})$, where $N_{e,\rm{ref}}$ is the number of reflected electrons and $N_{e,\rm{ups}}$ is the number of upstream electrons in the region over which spectra are calculated. This was obtained over the regions for which our spectra were calculated, and separating the low-energy and high-energy components at each timestep at the value of $\Gamma-1$ for which ${\rm d} (N_e(\Gamma-1))/ {\rm d}(\Gamma-1) = 0$. Electrons with $\Gamma - 1$ below and above this value were labeled as being upstream or reflected, respectively. The reflection rate as calculated by this method is shown in Fig. \ref{fig:refFrac}. As is consistent with earlier arguments, we see that simulations with lower values of $\thbn$ have larger electron reflection rates, with these reaching a high of $\sim 5\%$ for $\thbn=30^{\circ}$ and falling to only $\sim 0.05\%$ for $\thbn = 63^{\circ}$. We do not observe electron reflection for $\thbn=74.3^{\circ}$. We further note that the onset of electron reflection typically begins at later times for larger values of $\thbn$. This is consistent with them requiring a larger time to be pre-accelerated so that they have a large enough velocity to outrun the shock \citep{Amano2007}.

\subsubsection{Out-of-plane Simulations}

\begin{figure}
    \includegraphics[width=0.9\columnwidth]{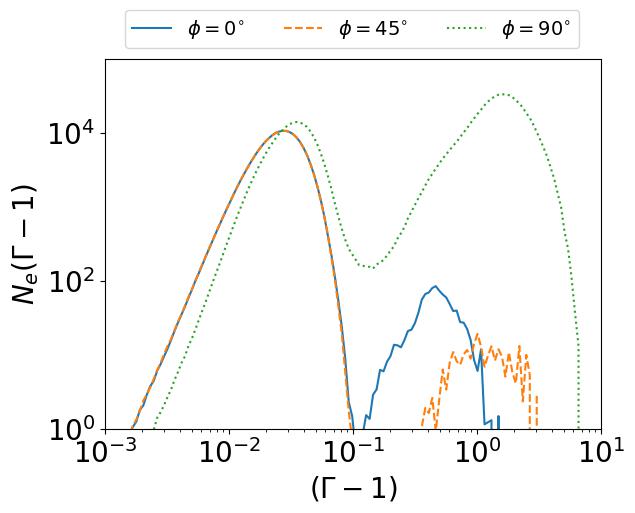}
    \caption{Spectra in the simulation frame for simulations with fixed $\thbn=63^{\circ}$ and varying the plane angle $\phi$. For the in-plane case, there is a very small overall difference in the reflection rate for $\phi=45^{\circ}$, but this change is significant when $\phi=90^{\circ}$. } \label{fig:specOop}
\end{figure}

Here we describe results from the simulations with a fixed $\thbn=63^{\circ}$ but varying $\phi$, such that $\phi = 0^{\circ}, 45^{\circ}, 90^{\circ}$. We focus on these rather than the simulations with $\thbn=74.3^{\circ}$ because for this value of $\thbn$, independent of $\phi$, we see no electron reflection. Fig. \ref{fig:specOop} shows the spectra from the simulations with $\thbn=63^{\circ}$ after $\Omega_{ci}t = 7.8$ in the simulation frame. We see significantly more reflection for the out-of-plane case where $\phi=90^{\circ}$ than for $\phi=0^{\circ}$ and $\phi=45^{\circ}$. The corresponding reflection rates were about $ 0.03\%$ for $\phi=0^{\circ}$ and $\phi=45^{\circ}$, but near $ 4\%$ for $\phi=90^{\circ}$. This immediately suggests an electron reflection mechanism which is more appropriately captured by out-of-plane simulations.

\subsection{Energy Content of EAWs}

In the wider context of astrophysical shocks, it is important to address the questions of how much kinetic energy these electrons carry back upstream, and whether it can drive any turbulence that affects the upstream plasma. The kinetic energy associated with a reflected electron is given by $E_K = (\Gamma - 1)m_e c^2$, and we have previously shown in Fig. \ref{fig:specTheta} that for larger $\thbn$ individual reflected electrons carry on average a higher $E_K$ but there are fewer of them in total.

\begin{figure}
    \centering
    \includegraphics[width=\columnwidth]{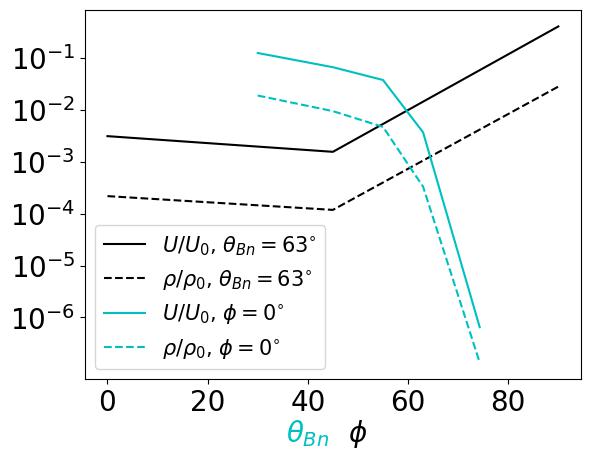}
    \caption{The solid lines show the normalized energy density and dashed lines the density ratio for in-plane (cyan, $\phi=0^{\circ}$) simulations with varying $\thbn$, whereas the black lines have fixed $\thbn=63^{\circ}$ and varying $\phi$. We see in general more energy is carried back upstream by reflected electrons for lower $\thbn$, and that this energy content is amplified significantly for an out-of-plane upstream magnetic field configuration.  }
    \label{fig:EdensComp}
\end{figure}

A more informative quantity is the energy density of reflected electrons, \mbox{$U_{\rm{ref}} = (m_e/2)R n_{\rm{ppc}} \overline{v_{\rm{ref}}}^2$}, which is defined using the reflection rate, $R$, and an average velocity of reflected electrons, $\overline{v_{\rm{ref}}}$. We normalize this quantity relative to the upstream electron energy density, $U_{\rm{ups}} = (m_e/2) n_{\rm{ppc}} v_{\rm{ups}}^2$, with the results shown for all simulations in Fig. \ref{fig:EdensComp}, along with the density ratios of the reflected and upstream electron beams. 

We see immediately that it is the higher reflection rate that dominates the total energy density of the reflected electron beam as opposed to the mean electron kinetic energy, as shocks with lower $\thbn$ have larger reflected energy densities, and their electron foreshocks comprise larger numbers of lower energy electrons relative to those at larger $\thbn$. We therefore expect that it is more likely for lower $\thbn$ shocks to have additional features in the electron foreshock region. Fig. \ref{fig:EdensComp} also shows shocks with $\thbn=63^{\circ}$ and varying plane-angle $\phi$ (see solid lines). Here, the energy density ratios are comparable for $\phi=0^{\circ}$ and $\phi=45^{\circ}$, but significantly higher for the out-of-plane case with $\phi=90^{\circ}$.

We now assess what happens to the energy content carried upstream by reflected electrons and focus on any turbulence it can drive in this region. This is especially important, as upstream electrons must interact with the electron foreshock before encountering the shock, so any mechanisms that can either lead to pre-acceleration or potentially compromise their transition to the shock downstream are necessary to understand the global picture of shock physics. 

Our in-plane simulations reveal the presence of EAWs far upstream of the shock, over the same spatial extent as where reflected electrons have propagated. Though these waves are also present in the out-of-plane shocks, they are less well captured as a result of the reflected electrons streaming on average along the upstream magnetic field lines, and hence wave vectors parallel to the streaming are not allowed \citep{Bohdan:2022}. For this reason, in this sub-section our analysis focuses on our in-plane results. 

To determine the energy content of the EAWs and demonstrate that they are powered by reflected electrons, we perform a 2-dimensional discrete Fourier transform (DFT) across the region of our simulation box where they are present. Following \citet{Vafin2018}, the dimensionless 2D Fourier amplitudes are calculated as,

\begin{align}
\begin{split}
    \widetilde{E}(k_{m,n}) &= \frac{e}{m_e c \omega_p} \frac{1}{N_x N_y} \cdot \\
    &\sum_{i=0}^{N_x-1} \sum_{j=0}^{N_y - 1} {\boldsymbol e_k} \cdot \boldsymbol{E}(x_i, y_j) \exp \left( 2 \pi i \left[ \frac{mi}{N_x} + \frac{nj}{N_y} \right] \right),
\end{split}
\end{align}
where $m$ and $n$ denote the respective indices of the $x$ and $y$ wave-vector components, and $\boldsymbol{e_k}$ is in the direction of the unit wave vector. Using Parseval's theorem, we can calculate the energy density associated with the electron acoustic waves by summing up $|\widetilde{E}(k_{m,n})|^2$ and dividing by the volume.

\begin{figure}
    \centering
    \includegraphics[width=\columnwidth]{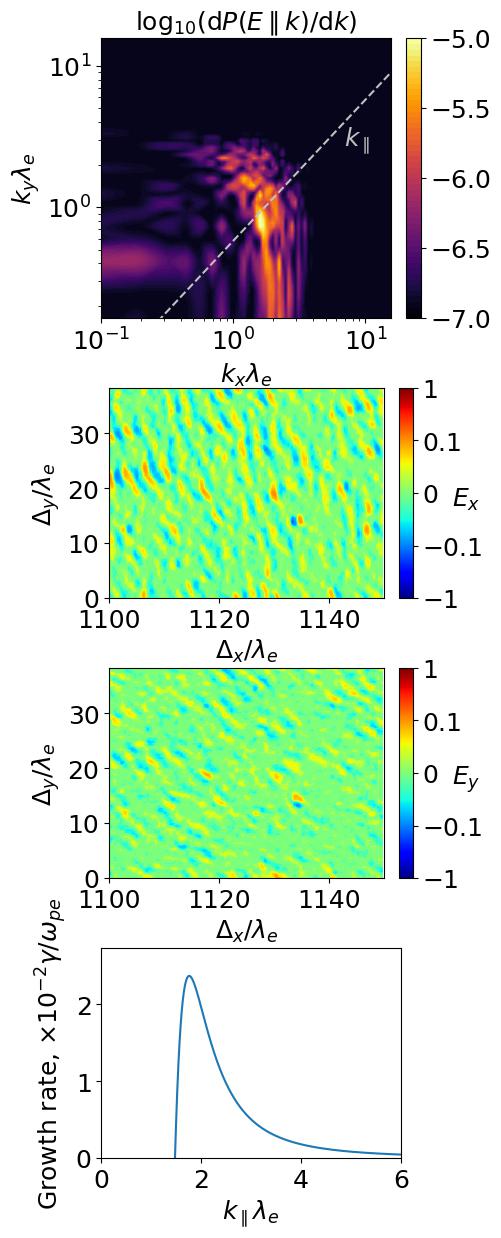}
    \caption{
    Top: Spectral power for $\mathbf{E} \parallel \mathbf{k}$ in units of $[c/(m_e c \omega_{p,e})]^2$ for the electric fields in the regions below (second: $E_x$, third: $E_y$).
    These maps show the EAWs for the $\thbn = 30^\circ$ simulation. Bottom: Growth rate of the electron-acoustic instability (eqn.~(\ref{eqnRelElAcoustic}), see text for parameters).}
    \label{fig:EFFT}
\end{figure}

The top panel of Fig. \ref{fig:EFFT} shows the DFT of the $x$- and $y$-components of the electric-field in an upstream region containing EAWs for $\thbn=30^{\circ}$. 
The wavelength of the fastest-growing mode here is consistent with the linear theory of the electron-acoustic instability. A Gaussian momentum distribution of reflected electrons excites electron-acoustic waves with a growth rate of \citep{Bohdan:2022}:
\begin{equation}
    \frac{\gamma(k_\parallel)}{\ompe}=-\sqrt{\frac{\pi}{2}}\frac{n_{\mathrm{ref}}}{n_{\mathrm{ups}}}\frac{m_e^2 \ompe^2}{p_{\mathrm{th}}^2}\ \frac{g(k_\parallel)}{k_\parallel^2} \exp \left(-\frac{g(k_\parallel)}{2}^2\right).
    \label{eqnRelElAcoustic}
\end{equation}
Here $n_{\mathrm{ref}}$ and $n_{\mathrm{ups}}$ denote the densities of reflected and incoming upstream electrons, respectively, and $$g(k)=(p_{\mathrm{res}}(k)-p_0)/p_{\mathrm{th}}$$
depends on the resonant momentum $p_{\mathrm{res}}(k)$, which follows from the resonance condition
\begin{equation}
    \frac{p_{\mathrm{res}}}{\left[m_e^2 c^2+p_{\mathrm{res}}^2\right]^{1/2}}=\frac{\ompe}{kc}.
\end{equation}
where the reflected electron beam has a momentum distribution centered around $p_0$ with a standard deviation $p_{\mathrm{th}}$.

We analytically determine the expected value of $k_{\parallel}$ by measuring the required quantities from our simulation data. Using the reflected electrons which occupy the region upstream of the shock shown in the central two panels of Fig. \ref{fig:EFFT}, we obtain values of $p_0=0.92~m_ec$, $p_{\mathrm{th}}=0.309~m_ec$ and $n_{\mathrm{ref}}=0.01~n_{\mathrm{ups}}$. For these parameters, eqn.~(\ref{eqnRelElAcoustic}) yields a peak growth rate of $0.024~\ompe$ at $k_{\parallel}~\lse = 1.76$, large enough for significant wave growth within small parts of the simulation time, $T\approx 5000\,\ompe^{-1}$.

The top panel of Fig. \ref{fig:EFFT} shows the Fourier transform of the electric field in this same region, allowing for a direct comparison of $k_{\parallel}$ obtained from our simulations to be made to the analytically expected value. The dashed silver line indicates the location of $k_{\parallel}$ which is in the direction of the reflected electron beam along the direction of the upstream magnetic field. From this, we calculate the observed scale of the electrostatic waves as $k_\parallel = k_x \cos \thbn + k_y \sin \thbn \sim 1.79$. The analytical value assumes the reflected electron beam follows a Gaussian distribution. As previously discussed,  electrons with higher energies are more likely to be reflected back upstream, hence the true distribution is not normally distributed in momentum space. Additionally, Fig. \ref{fig:refFrac} shows that we observe temporal fluctuations in the electron reflection rate, which from eqn. \ref{eqnRelElAcoustic} we expect to manifest proportionally in the growth rate of the EAWs, meaning this quantity will not necessarily be constant across a given upstream region of our simulation box. Despite these discrepancies between theoretical calculations and simulation results we nevertheless achieve a very good match in $k_{\parallel}$, strongly supporting the notion that these waves are indeed generated by the electron acoustic instability.

Furthermore, we note that the most energetic reflected electrons are found at the leading edge of the foreshock. This time-of-flight effect arises naturally as these more energetic and therefore faster electrons are more capable of outrunning the shock. That they have larger streaming speeds means that the associated electron acoustic waves have larger wavelengths further out from the shock, while the lower number density of these electrons reduces the total power contained in these waves.

The total energy contained in the EAWs across our the upstream region of our simulations is computed by using Parseval's theorem applied to a window which we slide upstream in our simulation domain. This window is square in shape with size  $40 \lambda_{\rm{se}} \times 40 \lambda_{\rm{se}}$ which ensures the electron acoustic waves are appropriately sampled. As we are not interested in the downstream region here, the first window begins at the same point upstream as was used to define our spectra calculation. Likewise, we utilize the same methods to measure spectra and quantify the kinetic energies of the reflected and upstream components from them.

\begin{figure}
    \includegraphics[width=\columnwidth]{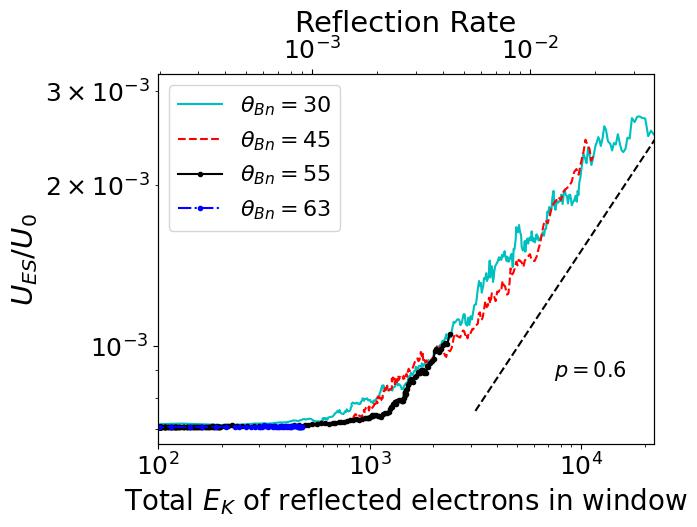}
    \caption{The energy density of the electrostatic waves in the electron foreshock, normalized to the bulk kinetic energy density of upstream electrons, plotted as a function of the total kinetic energy carried by reflected electrons. The strong correlation present in all simulations here supports the view that reflected electrons drive these waves.
    }
    \label{fig:UesVsR}
\end{figure}

Fig. \ref{fig:UesVsR} plots the normalized energy density contained in these electrostatic waves as calculated from summing up the Fourier power in each $40 \lambda_{\rm{se}} \times 40 \lambda_{\rm{se}}$ window as a function of the kinetic energy of the contained reflected electrons, again normalized to that upstream. It is clear that when an extended foreshock is present, a strong correlation exists between the two, further indicating that these waves are indeed driven by reflected electrons. Here, the floor at around $U_{ES}/U_0 \sim 7\times10^{-4}$ is likely a result of electrostatic noise, making a threshold reflection rate for the generation of EAWs difficult to ascertain.

\subsection{Electron Reflection Mechanisms} \label{sec:initRef}

As EAWs are excited by a velocity difference between cold incoming and hot reflected electrons, it is important what causes the reflection of the hot component. It has long been established that ions that are reflected by the electrostatic potential at the leading edge of the shock can generate electrostatic waves upon their interaction with incoming electrons \citep{Hoshino2002}. Note that electrons are oppositely charged to ions, so cannot be reflected by the shock potential, and instead would be accelerated downstream. The requirement to excite these so-called Buneman waves is that the relative velocity difference between the two particle species, $\Delta v$, must exceed the thermal speed of the upstream electrons \citep{Buneman1958}. They have a wavelength of \citep[e.g.][]{2009PhPl...16j2901A}, $\lambda_B = (2\pi/c) \Delta v \lse \approx (2\pi/c) \frac{3}{2}  v_{\rm{sh}} \lse$ where $\lse$ is the electron skin length. 

\begin{figure*}
    \centering
    \includegraphics[width=\textwidth]{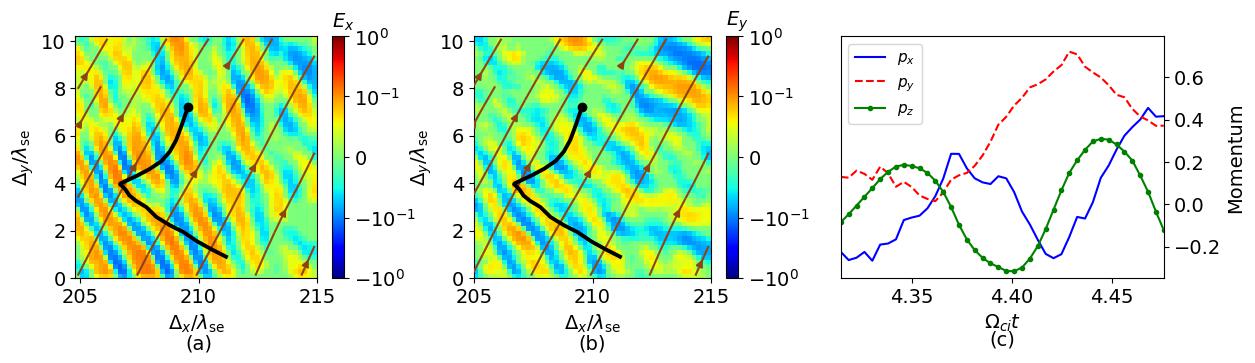}
    \caption{
    The color maps in panels (a) and (b) represent the $x$- and $y$- components of the electric field encountered by the electron, whose path is shown by the black line (the black circle gives the final location) in units of the electron skin length, $\lambda_{\rm{se}}$. We see a clear interaction between the electron and the electrostatic wave, resulting in a net gain of momentum in the $x$- and $y$- directions (panel c). If this energy gain is sufficient, it is possible for the electron to outrun the shock and be reflected. The plot displays quantities in the simulation frame for $\thbn=55^{\circ}$, with the direction of the upstream magnetic field indicated by the brown arrows. 
    }
    \label{fig:parRef}
\end{figure*}

\begin{figure}
    \centering
    \includegraphics[width=\columnwidth]{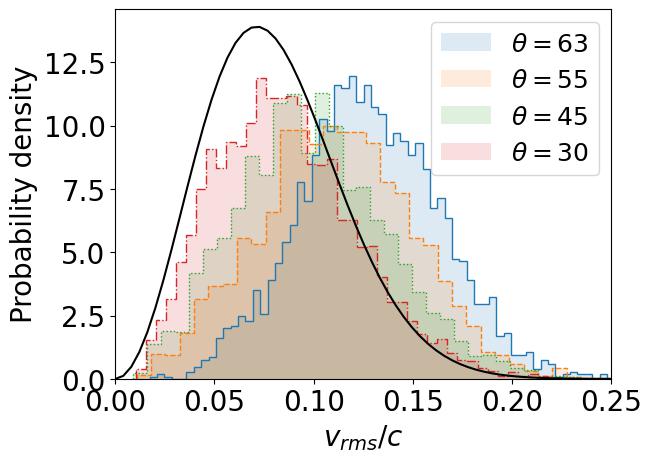}
    \caption{Initial velocity distribution of to-be reflected electrons in the upstream rest frame. The quantities here are averages over the first gyro-period after injection into the simulation upstream. Black line: initial thermal distribution of upstream electrons with arbitrary normalization. We can see that the sub-set of electrons that are reflected are initially more energetic than the most probable thermal velocity. }
    \label{fig:initDistr}
\end{figure}

Fig. \ref{fig:parRef} shows the early-time reflection of a typical electron within our simulation. Its trajectory is shown by the black line, with the color maps illustrating the strength of the $x$- (leftmost) and $y$- (center) components of the electric field. Here, the electron, which is initially traveling with a velocity x-component directed from right to left ($-\boldsymbol{x}$ direction) interacts with this Buneman wave and turns back. We note also from the trajectory that this electron has a relatively large $y$-momentum, as without this it would simply move with the bulk flow from right to left. This detail is important, as a comparatively large transverse momentum allows the electron to relatively freely enter the potential well associated with the electrostatic wave. Fig. \ref{fig:initDistr} confirms that the more energetic electrons injected into the upstream plasma are more favorably reflected in this manner. Once trapped, the electron can undergo shock-surfing acceleration where it is accelerated by the motional electric field, or be accelerated by the electrostatic field itself \citep[see appendix of][]{Bohdan2019a}. To escape the potential well, the trapping force ($q E_{BI}$, where $E_{BI}$ is the electric field associated with a Buneman wave) must be exceeded by the Lorentz force ($q_e v_e B_0/c$) \citep{Hoshino2002,Matsumoto2012}. 

Importantly, the involvement of Buneman waves in the initial reflection process is supported by the fact our simulations show a larger electron reflection rate when the plane-angle $\phi=90^{\circ}$. For an out-of-plane magnetic field configuration, the energization of electrons by Buneman waves relative to the in-plane case is more efficient \citep[][]{Bohdan2017,Bohdan2019a}, thus the probability that they gain enough energy to outrun the shock and escape upstream is increased. This is because the gyromotions of the ions driving the Buneman waves are appropriately captured in the simulation plane for $\phi=90^{\circ}$, but this is not the case for other angles. Accordingly, electron shock surfing on Buneman waves becomes more efficient, though this alone cannot explain why they are reflected.

To address this, we consider the direction in which work is done on individual electrons. This can be separated into components either in the perpendicular or parallel direction to the local magnetic field. One can visualize the former as bringing about a change in the gyroradius, $r_{ge} = p_{\bot}/(q_e B)$ while solely increasing the parallel velocity leads to a decrease in pitch angle, $\alpha = \arctan \left(v_{\bot}/v_{\parallel}\right)$.

Perpendicular work naturally arises from acceleration due to the motional upstream electric field, defined as $\boldsymbol{E} = - \boldsymbol{v_{\rm ups}} \times \boldsymbol{B_0}$, thus is oriented perpendicularly to the initial upstream magnetic field in the electron foreshock that we are interested in here. It essentially constitutes gyration in the rest frame of the magnetic field. One such mechanism that increases perpendicular momentum is shock surfing acceleration \citep[e.g.][]{2003P&SS...51..665S,2009ApJ...690..244A}, which relies on the formation of Buneman waves described earlier. Upstream electrons can then become trapped in the potential wells of these waves, and are accelerated perpendicularly by the motional electric field \citep[e.g.][]{2000ApJ...543L..67S,Hoshino2002}.

Conversely, for electrons to be accelerated parallel to the local magnetic field, the plasma must be somehow disturbed, otherwise $\boldsymbol{E}$ and $\boldsymbol{B}$ remain perpendicular to each other. Supposing, hypothetically, an electron was instantaneously reflected back upstream with the transformation $v_x \rightarrow -v_x$ and an oblique upstream magnetic field orientation described by $B = (B_0\cos\thbn, B_0\sin \thbn) $. Here, we can see intuitively that the sign of $\ppar$ reverses, while we expect $|\pperp|$ to remain constant. An example of this is magnetic mirroring \citep[e.g.][]{Ball2001}, in which a particle encountering a shock is either reflected back upstream or transmitted downstream, depending on the pitch angle. A signature property of this process is the conservation of the first adiabatic invariant, $\mu$, defined by,

\begin{equation}
    \mu = \frac{m_e v_{\bot}^2}{2B} = \frac{m_e v^2 \sin^2 \alpha}{2B},
\end{equation}
for pitch angle $\alpha$. For $\mu$ to remain constant, the kinetic energy of the particle must scale with the magnetic field strength. The condition for reflection can be derived by Lorentz transforming into the de-Hoffmann-Teller frame in which the fluid velocities are parallel to the magnetic field lines on both side of the shock \citep{1950PhRv...80..692D}. As there are no electric fields \citep[e.g.][]{Toptygin1980} and noting that $v^2$ is the same before and after reflection, we find $\sin^2 \alpha_u / B_u = \sin^2 \alpha_d /B_d $  for a particle crossing from upstream (subscript $u$) to downstream (subscript $d$). Following \citet{Ball2001}, we see that if a particle had $\sin^2 \alpha_u > B_u/B_d$, it would have a undefined pitch angle downstream. Instead it is reflected, with the reflection condition in this frame being $\alpha_u > \alpha_c$ for $ \sin^2 \alpha_c = B_u/B_{\mathrm{max}}$, with the subscript $c$ here denoting the critical value and $B_{\mathrm{max}}$ the peak of the magnetic field along the particle trajectory. 

\begin{figure}
    \centering
    \includegraphics[width=\columnwidth]{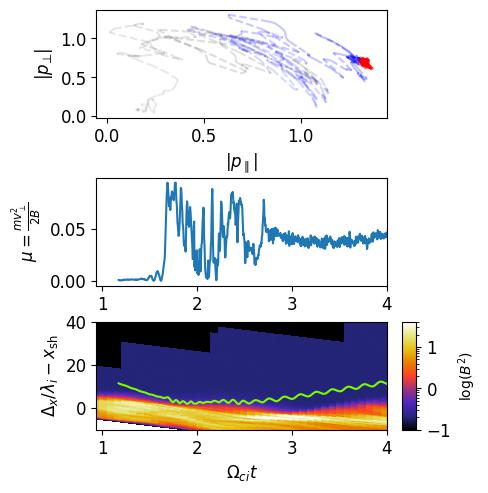}
    \caption{For one of the first reflected particles in the simulation with $\thbn = 63^{\circ}$, the top panel plots the perpendicular momentum against that parallel to the local magnetic field. Here, the black section represents the interval $\Omega_{ci}t<2$, the blue $2 \leq \Omega_{ci}t<3$ and the red $3 \leq \Omega_{ci}t<4$. The initial sharp rise in $p_{\bot}$ coincides with the rapid change in magnetic moment, $\mu$ (second panel, measured in the upstream rest frame), when the electron is shock surfing at the leading edge of the shock (lower panel). Here, $x_{\rm{sh}}$ is defined assuming the analytical shock velocity in the simulation frame of $v_{\rm{sh}}' \approx 0.067c$. After surfing for a while, the electron is reflected, during which $\mu$ does not change so dramatically.}
    \label{fig:SSA_then_ref}
\end{figure}

Fig. \ref{fig:SSA_then_ref} shows some of the properties of one of the earliest reflected electrons in the simulation with $\thbn=63^{\circ}$. The green line in the lower panel shows the trajectory of the electron, with its location along the $x$-axis of the simulation box plotted against time. At around $\Omega_{\rm{ci}}t \sim 1.75$, the electron encounters Buneman waves in the shock foot, and begins to shock-surf, which is indicated both by the sudden increase of the first adiabatic invariant and the large initial increase in $|p_{\bot}|$ relative to $|p_{\parallel}|$. 

At around $\Omega_{\rm{ci}}t \sim 2.5$, the electron is reflected from the shock with the first adiabatic moment remaining roughly constant, indicating magnetic mirroring. This two-step reflection process is that described by \citet{Amano2007}, who showed that electrons are initially energized in the perpendicular direction by shock surfing acceleration in the shock foot, and are then further accelerated and reflected by magnetic mirroring (or shock drift acceleration). The initial perpendicular acceleration from SSA helps to maintain a large pitch angle, which results in a higher probability of the electron fulfilling the requirements for reflection instead of transmission through the shock.

Once enough electrons have been reflected in this manner, their velocity difference relative to the upstream electrons can excite the aforementioned electron acoustic instability and generate EAWs in the electron foreshock.

\section{EAW-electron interactions}

\begin{figure}
    \centering
    \includegraphics[width=\columnwidth]{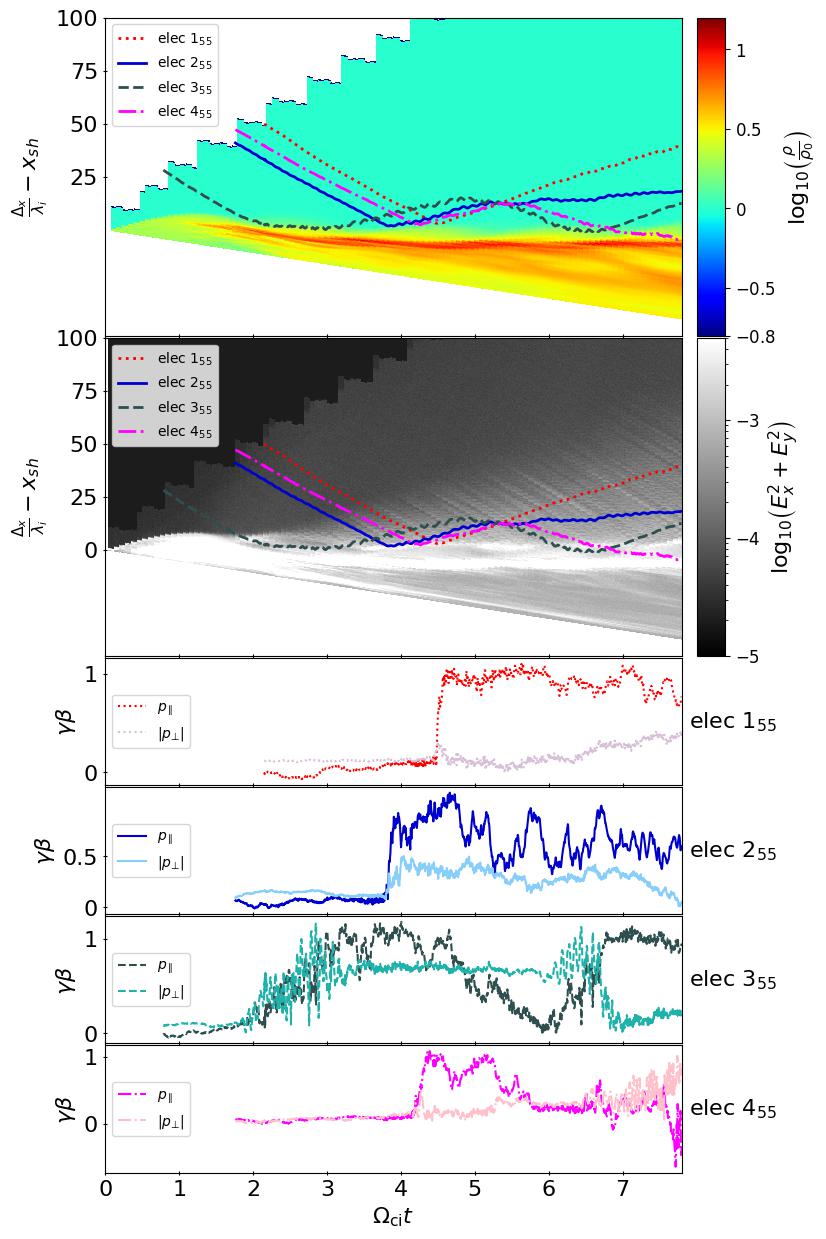}
    \caption{For $\thbn=55^{\circ}$. The top panel provides as a color map the time evolution of the normalized electron density profile along the simulation box (in ion skin lengths, $\lambda_i$) relative to $x_{\rm{sh}}$ which is assumed to move at a constant speed of $v_{\rm{sh}} \approx0.067c$. The second panel does the same, but averages $E_x^2 + E_y^2$ to show the electrostatic wave power in the electron foreshock. These figures have trajectories of selected electrons superimposed, with the colors and line-styles corresponding to the lower figures. These lower panels display the momenta of selected electrons parallel and perpendicular to the magnetic field in the upstream rest frame, and defines the label of each for discussion in the main text.}
    \label{fig:th55refElecs}
\end{figure}

Having discussed the general properties of the electron foreshock in the previous section, we discuss what happens upstream following the production of EAWs. To achieve this, we take advantage of the particle tracing feature in our code, periodically recording the locations, velocity and local field components for each particle. In this section, we once again focus on our in-plane simulations for the reason that by definition the electron acoustic waves generated by the reflected electrons reside in the simulation plane, so are well captured when $\phi=0$.

For each simulation, we trace a sample of 10,000 electrons. To ensure that we can trace the entire evolutionary history of these particles beginning with their injection into the simulation upstream until after their reflection, we select particles that already reside in the foreshock (so have already been reflected) at $\Omega_{{\rm ci}} t = 5$. Our code can then be re-run at earlier timesteps and record data for these particles from their injection into the upstream up until the end of the simulation. We select foreshock electrons by imposing a number of criteria. Firstly, we select particles more than $5\lambda_i$ further upstream of the shock foot location, $x_{\rm{foot}}$. This prevents electrons that are currently undergoing shock surfing acceleration, but may end up downstream, from being selected. Secondly, we ensure that the magnitude of the $x$-component of their velocity in the upstream rest frame exceeds the shock velocity, $v_{e,x} > u_{\rm{sh}}$. This guarantees they can, unless they are decelerated, outrun the shock and that at some point they must have been reflected. Finally, we also impose the condition that their velocity $y$-component must be positive, $v_{y} > 0$, as reflected electrons should be tied to the upstream magnetic field, which in these simulations always lies in the first quadrant of the $x$-$y$ plane. The individual and unique IDs of all electrons fulfilling these criteria are output, and a random selection is made to ensure different behaviors are captured and limit any selection bias. 

\begin{figure}
    \centering
    \includegraphics[width=\columnwidth]{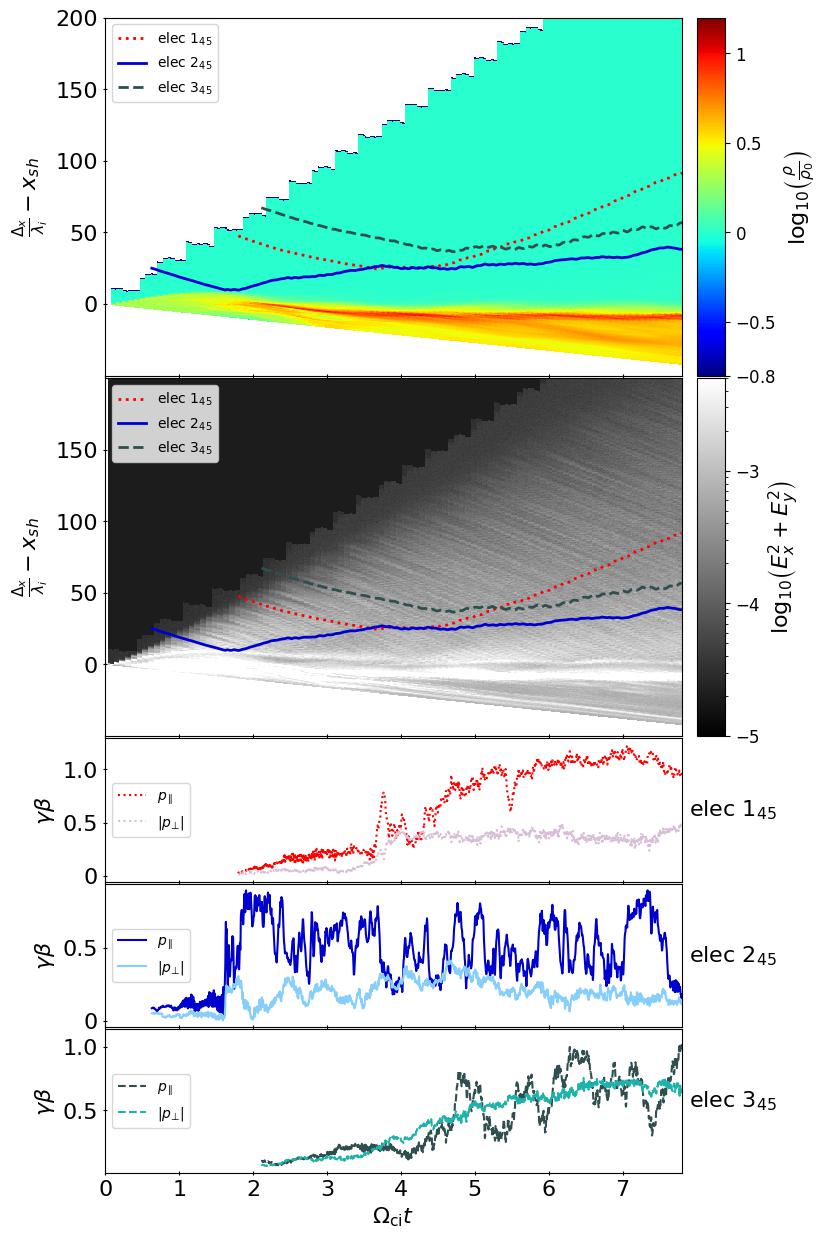}
    \caption{As Fig. \ref{fig:th55refElecs} but for the simulation with $\thbn = 45^{\circ}$. The colors of the selected particle trajectories match those in the lower momentum sub-plots.}
    \label{fig:th45refElecs}
\end{figure}

We illustrate our main results in Figures \ref{fig:th55refElecs} and \ref{fig:th45refElecs}, which summarize the results for the simulations with $\thbn = 55^{\circ}$ and $\thbn = 45^{\circ}$, respectively. In both figures, the top panel shows the density profile in the simulation box averaged over the transverse direction as a color map that is plotted as a function of time and the distance to the shock, $x_{\rm{sh}}$, which is calculated assuming it moves with its analytical velocity value of $v_{\rm{sh}}' \approx 0.067c$ in the simulation frame. To show the effect of the electrostatic waves present in $x$-$y$ (simulation) plane of the electron foreshock, the second panel shows $\log_{10} (E_x^2 + E_y^2)$. In both figures we see a clear shock transition, which is sharper in the density profile for $\thbn = 45^{\circ}$ relative to $\thbn = 55^{\circ}$, and likewise stronger in $\log_{10} (E_x^2 + E_y^2)$, as a consequence of the higher reflection rate and larger total energy content carried by the reflected electrons (see Fig. \ref{fig:EdensComp}). 

These upper panels have the trajectories of selected reflected electrons superimposed. Together they encapsulate the range of physical behaviors that electrons experience when traveling through the electron foreshock and encountering the EAWs. The lower panels illustrate the momentum history of the selected electrons, separated into the components parallel, $p_{\parallel}$, and perpendicular, $p_{\bot}$, to the upstream magnetic field. The color and linestyle of each electron trajectory in the upper panel matches the lower panel corresponding to it, with each electron assigned a label to the right of each lower panel. It is immediately apparent that a wide range of possible trajectories, with accompanying changes in particle momentum are possible. The physical origin of each case are discussed in turn.

We see from Figures \ref{fig:th55refElecs} and \ref{fig:th45refElecs} that some of the selected electrons exhibit sharp increases in $\ppar$ while $\pperp$ remains approximately constant, which is indicative of them being reflected by magnetic mirroring. This effect is most profound for electrons $1_{55}$ and $4_{55}$, between $\Omega_{ci}t = 4 - 5$, and is also apparent for electron $2_{45}$ at $\Omega_{ci}t \approx 1.5$. While the momentum history of these particles is comparable around the time of mirroring, they initially also undergo some shock surfing, experiencing the reflection mechanism described in \citet{Amano2007} (cf. Section \ref{sec:initRef}). As already discussed, we find that this two-step acceleration plays a role at early times and helps to reflect the initial foreshock electrons. 

Figs. \ref{fig:th55refElecs} and \ref{fig:th45refElecs} reveal further changes in the particle momenta after reflection. This effect is particularly noticeable in the parallel direction, as such the work required cannot be done by the motional electric field and instead comes from the EAWs in the electron foreshock oriented, on average, along the direction of the upstream magnetic field (because reflected electrons travel, on average, in this direction). We note that the values of $\ppar$ and $\pperp$ typically do not remain constant after the initial reflection, ultimately leading a range of possible outcomes within the $\Omega_{\rm{ci}}t = 7.8 $ timespan of our simulations. We attribute this behavior as electrons becoming trapped by an electrostatic wave and be convected either back towards or further away from the shock. For electrons $1_{55}$, $2_{55}$, and $2_{45}$ these associated momentum changes appear to be relatively inconsequential, with the parallel particle momenta jumping around a mean value with respect to time. For electron $2_{55}$, a reduction in $p_{\parallel}$ can be seen at late times in the top two panels of Fig. \ref{fig:th55refElecs}, with the gradient of the blue line decreasing, allowing the relative distance between this electron and the shock to become smaller. Electron $3_{55}$ initially undergoes steady parallel acceleration before shock surfing and ultimately obtaining enough energy to be reflected. During this stage, the parallel component of its velocity matches the phase velocity of the EAWs \citep{Bohdan:2022}, supporting the notion that it is Landau trapped. Eventually, while this electron is spatially separated from the shock at $\Omega_{\rm{ci}}t \approx 3-6$, it is advected back towards the it, losing $\ppar$ at $\Omega_{\rm{ci}}t \approx 5$. The relative percentage momentum change is significantly smaller for faster moving electrons such as electron $1_{55}$, which can be explained by limited Landau damping as the velocity is not in resonance with the phase velocity of the EAWs.

Multiple reflection events are not the only possible outcomes for foreshock-scattered electrons. Electron $4_{55}$ ends up in the downstream region some four gyrotimes later than the initial reflection at  $\Omega_{\rm{ci}}t \approx 4$. Here, the electron re-interacts with the shock with significantly more energy than it was injected into the upstream with, which can result in further and more efficient acceleration via stochastic Fermi acceleration \citep{Bohdan2019b} or stochastic shock-drift acceleration \citep{Matsumoto2017,Katou2019}. This is important, as the extra energy they gain from the earlier acceleration processes could allow them to more easily pass into the downstream than compared to if they had thermal energies. 

Another interesting feature of Fig. \ref{fig:th45refElecs} are the differing trajectories of electrons $1_{45}$ and $3_{45}$. The location history of these electrons shows that they do not get within $50 \lambda_i$ of the shock discontinuity, yet are still turned back upstream. We note that trajectories like this did also occur for $\thbn = 55^{\circ}$, yet, due to the less energetic foreshock, were somewhat rarer, occurring for only 70 of our 10,000 traced electrons in comparison to $\sim 15\%$ for $\thbn=45^{\circ}$. For magnetic mirroring to be the mechanism responsible for reflection, we need a converging magnetic field, which is not present in the electrostatic foreshock. 

\begin{figure}
    \includegraphics[width=\columnwidth]{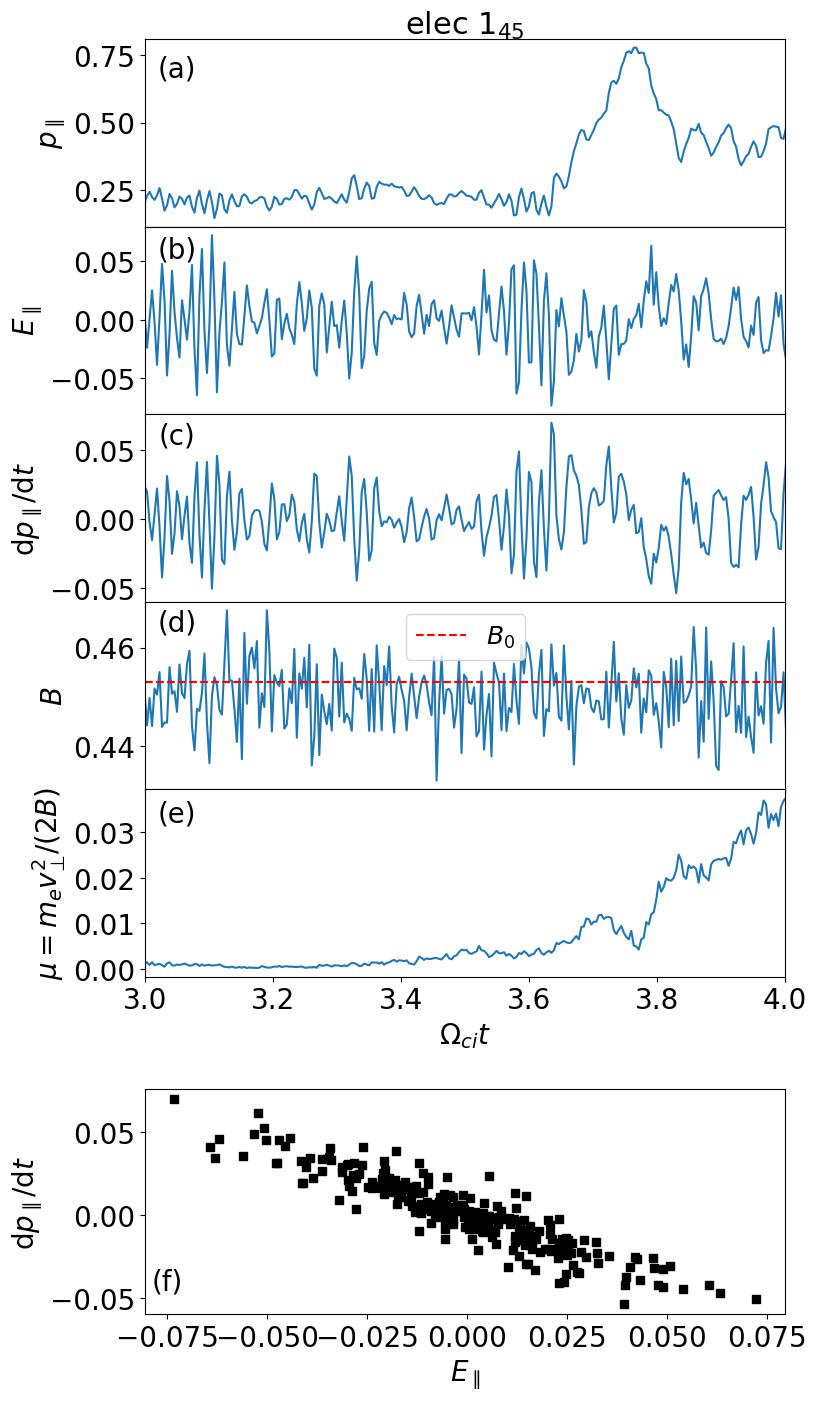}
    \caption{ Illustration of the local field dependencies as a function of time for electron $1_{45}$ as measured in the upstream rest frame. The figure plots the time interval from Fig. \ref{fig:th45refElecs} with the first substantial increase in $p_{\parallel}$ (panel a). Subsequent panels show the time evoltion of (b) $E_{\parallel}$, (c) ${\rm d}p_{\parallel}/{\rm d}t$, $B$ and $\mu$. Panel (f) shows that see a strong negative correlation between $E_{\parallel}$ and ${\rm d}p_{\parallel}/{\rm d}t$. A similar plot with $|B|$ plotted on the $x$-axis shows no trend (not plotted), reinforcing the inference that this electron is turned back upstream by the electric field associated with the EAWs. 
    } \label{fig:elec1_45}
\end{figure}

We further demonstrate the ability of the EAWs to turn electrons back upstream in Fig. \ref{fig:elec1_45}, which shows the time evolution of $p_{\parallel}$ between $\Omega_{\rm{ci}}t = 3$ and $\Omega_{\rm{ci}}t = 4$ in panel (a). This interval was selected as it is during this period that the electron is turned away from the shock, despite always residing in the upstream. The quantities here are measured in the upstream rest frame. The accompanying panels display the corresponding time evolution of (b) $E_{\parallel}$, (c) ${\rm d}p_{\parallel}/{\rm d}t$, (d) $B$ and (e) $\mu$. Panel (f) illustrates a strong correlation between  ${\rm d}p_{\parallel}/{\rm d}t$ and $E_{\parallel}$, suggesting that this field component, which is associated with the EAWs, is responsible for the change in  $p_{\parallel}$. This is further reinforced by the fact that the local magnetic field experienced by the electron is consistent with the upstream value and shows no correlation with ${\rm d}p_{\parallel}/{\rm d}t$, and that the first adiabatic invariant is not constant and in fact steadily increases during the reflection, ruling out magnetic mirroring. 

We propose the following overall picture: initially, electrons  undergo shock surfing acceleration on Buneman waves and are reflected via magnetic mirroring. If enough electrons are reflected with a large enough total energy content, we begin to see electron acoustic waves traveling into the shock foot, with the overall energy density of these waves increasing with the reflection rate, hence is higher closer to the shock, and at later times once the reflection rate begins to saturate. These waves have a $k$-vector approximately aligned with the upstream magnetic field, and can do work parallel to it. On average, electrons encountering these waves will experience a small net force in the direction of the upstream magnetic field, which leads to their momenta being aligned increasingly towards it instead of in the direction of the bulk flow (towards the shock). Analogous to Buneman waves, it is possible for upstream electrons to become trapped in the potential wells of the EAWs if the potential difference is sufficiently large. Acceleration by the motional electric field leads to an increase in $\pperp$, as is the case for electron $3_{45}$. Furthermore, they can be accelerated by the on-average parallel electric field. Subsequent interactions between these electrons and the EAWs can lead to them being turned further away from the shock, or convected back towards it, where they may be caught up with the shock and again interact with it. 

\begin{figure}
    \centering
    \includegraphics[width=\columnwidth]{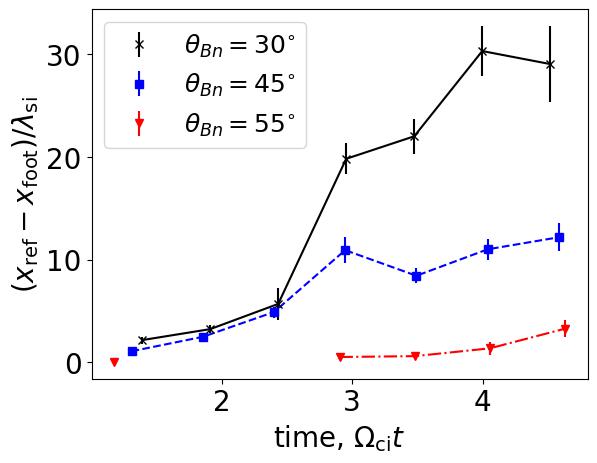}
    \caption{The mean electron reflection distance from the shock foot position as a function of time for trace particles. We see that this value increases with time as more electrons are reflected and power electron acoustic waves in the foreshock. }
    \label{fig:xrefVt}
\end{figure}

As with Buneman waves, the EAWs can accelerate upstream electrons and even reflect them back upstream. The reflection of electrons in this manner will again help to drive further upstream turbulence, and contribute to further electron reflection and pre-acceleration. This is indicated in Fig. \ref{fig:xrefVt}, which illustrates that the mean reflection location of an electron with respect to $x_{\rm{foot}}$ increases as a function of time. Here, for each reflected electron, the reflection point was defined as the minimum distance from $x_{\rm{foot}}$ and only includes the first reflection point for electrons with multiple shock interactions. 

By using our tracing data in combination with the definition of the shock foot, $x_{\rm{foot}}$, we can estimate how often reflected electrons interact with the shock. We define a single shock interaction when an electron crosses from $x > x_{\rm{foot}}$ to $x \leq x_{\rm{foot}}$. In general, for all simulations, $70\% - 80\%$ of all reflected electrons have a single interaction, indicating the reflection mechanism is that described in \citet{Amano2007}. The percentage of all reflected electrons with 2 or more interactions was $\sim 30\%$ for $\thbn = 63^{\circ}$, decreasing to $\sim 9\%$ for $\thbn = 45^{\circ}$. This is because at higher obliquities, electrons require a larger speed to escape the shock and therefore need more pre-acceleration, such as more cycles of SDA. In comparison, the higher EAW energy density contained in EAWs means that the percentage of reflected electrons  turned back by interacting with them was higher at smaller $\thbn$, with $\sim 15\%$ for $\thbn = 45^{\circ}$ and $\sim 0.7\%$ for $\thbn = 55^{\circ}$ and negligible for $\thbn \geq 55^{\circ}$. For $\thbn = 30^{\circ}$, the shock takes longer to form \citep[e.g.][]{Gingell:2017} and is in fact dominated by an ion foreshock close to the foot, making a comparison challenging and beyond the scope of this work. 

In \citet{Bohdan:2022}, a similar simulation to those here is presented, with the main differences a wider simulation box, a total run-time of $\Omega_{\rm ci}t = 50.4$, and an initial upstream magnetic field configuration characterized by $\thbn=60^{\circ}$ and $\phi=90^{\circ}$. By integrating the energy density of shock-reflected electrons up to a threshold value where EAWs can be detected, it is estimated that the EAW region will saturate at approximately $\Omega_{\rm ci}t \sim 270$. Furthermore, similarly to our $\thbn=30^{\circ}$ simulation (although with a later onset), substantial ion reflection begins in the  \citet{Bohdan:2022} simulation at around $\Omega_{\rm ci}t \sim 10$, which causes Whistler waves to form immediately ahead of the shock transition. We expect such features to occur in the simulations presented here at later times, and aim to investigate pre-acceleration associated with the Whistlers in future work.

\section{Conclusions}

In this work, we have performed PIC simulations which have revealed the presence of electrostatic waves, identified as electron acoustic waves, which are generated when reflected electrons excite the electron acoustic instability. We have investigated the properties of these waves and their influence on the upstream electrons, with the key findings of our work summarized below:

\begin{itemize}
    \item PIC simulations were completed for the range of angles between the shock normal and upstream magnetic field of $\thbn = \left[ 30, 45, 55, 63, 74.3 \right]\ \mathrm{degrees}$. This allowed for a detailed study of the rates of incident electrons that are reflected back upstream and assess their effect on the upstream plasma.
    \item A higher percentage of incoming electrons are reflected back upstream for smaller $\thbn$. About $ 5\%$ of electrons were reflected for $\thbn=30^\circ$, but only around $0.05 \%$ for $\thbn=63^\circ$.
    \item The initial reflection mechanism involves shock surfing acceleration on Buneman waves followed by magnetic mirroring, and is described in \citet{Amano2007}. Our results show a higher reflection rate for out-of-plane simulations, verifying the role of Buneman waves in the initial electron reflection.
    \item Although fewer electrons are reflected at higher $\thbn$, the typical peak energy of these electrons is higher. This is because they require velocities $v_e \cos \thbn > v_{\rm{sh}}$. Despite this, the greater number of electrons reflected in shocks with smaller $\thbn$ means overall there is a greater energy density in reflected electrons for these angles.
    \item Reflected electrons lead to the production of electron acoustic waves in the electron foreshock on spatial scales comparable to Buneman waves.
    \item These electron acoustic waves are themselves capable of accelerating and reflecting upstream electrons further away from the shock, preventing them from reaching it and compromising injection for $\sim 1\%$ of upstream electrons for $\thbn \leq 45^{\circ}$.

\end{itemize}

\acknowledgments

M.P. acknowledges support by DFG through grant PO 1508/10-1. The numerical simulations were conducted on resources provided by the North-German Supercomputing Alliance (HLRN) under project bbp00033.

\bibliographystyle{apj}
\bibliography{Bibliography}

\end{document}